# Similarity analysis of the streamer zone of Blue Jets

N.A. Popov[1], M.N. Shneider[2] and G. M. Milikh[3]

[1]Moscow State University, Moscow, Russia

[2]Department of mechanical and aerospace engineering Princeton University, Princeton, NJ, USA

[3]Astronomy Department University of Maryland, College Park, MD, USA

## Abstract

Multiple observations of Blue Jets (BJ)) show that BJ emits a fan of streamers similar to a laboratory leader. Moreover,in the exponential atmosphere those long streamers grow preferentially upward, producing a narrow coneconfined by the aperture angle. It was also noticed that BJ are similar to the streamer zone of a leader (streamer corona) and the modeling studies based on the streamers fractal structure were conducted. Objective of this paper is to study the fractal dimension of the bunch of streamer channels emitted by BJ, at different altitude and under the varying reduced electric field. This similarity analysis has been done in three steps: First we described the dendritic structure of streamers in corona discharge applying the fractal theory. Then using this model and the data from existing laboratory experiments we obtained the fractal dimension of the branching streamer channels. Finally the model was validated by the observations of BJ available from the literature.

## 1. Introduction

About two decades ago, researchers discovered upward-propagating collimated flashes of light originating above thunderstorms [*Wescott et al.,* 1995]. Due to their distinctive, principally, blue color, they were termed "blue jets" (BJ). They resemble tall trees with quasi-vertical trunk and filamentary branches. [*Pasko et al.*, 2002] discovered the so-called gigantic blue jets (GBJ), propagating into the lower ionosphere. Number of GBJ observations were made since [*Su et al.,* 2003; *Kuo et al.,* 2009; *Cummer et al.,* 2009; *Lu et al.,* 2011; *Chou et al.,* 2011].

Multiple observations of BJ show that they emits a streamers similar the streamers of corona discharge of a laboratory leader. Streamers in corona discharge are self-similar plasma structures termed fractals [*Wiesmann et al.,* 1986; *Pasko et al.*, 2000; *Popov,* 2002]. According to the current understanding the streamer propagates due to the formation of electron avalanches in the strong electric field $E$ of the space charge near the streamer tip. The propagation direction of those avalanches is determined by the shape of the streamer's electric field. Furthermore the probability of streamer propagation in a given direction is determined by the spatial distribution of the electric field near the streamer tip.

The dendritic structure of streamers in corona discharge was treated by means of the fractal theory [*Satpathy, 1986*], which main parameter is the fractal dimension $D$. Note that the fractal dimension $D$ of the plasma structure is determined by the power index $\gamma$, which depends on the reduced electric field value near the streamer tip [*Popov, 2002*]. [*Pasko et al., 2000*] numerically simulated the streamer corona of a positive leader as a fractal process. They suggested that the power index $\gamma$, that links the probabilities a random choice with electric field value, is $\gamma = 1$ and obtained the fractal dimension of the dendritic structure $D = 1.37 \pm 0.13$.

The objective of this paper is to apply the fractal model of streamers in the corona discharge to study formation and propagation of blue jets and gigantic blue jets. The model is validated by comparison with the existing optical observations of blue jets and gigantic blue jets.

## 2. Fractal dimension of a bunch of multiple streamers

Widely used model [*Niemeyer et al., 1984; Wiesmann et al., 1986*] describes the spatial structure of branching channels. In this model, the plasma channels propagate toward the highest electric field, provided that the field exceeds the threshold value $E_{cr}$. The spatial distribution of the electric field is determined from solution of the Laplas equation for the potential having the boundary conditions which consider the conducting channels. All these channels are taking as equal, while the field inside the channels is constant. The probability of the streamer propagation in z-direction $P_z$ is determined by the field distribution near the streamer head.

$$P_z \propto E_z^\gamma - E_{cr}^\gamma \qquad (1)$$

where $E_z$ is the field in the direction z, $\gamma$ is the power index.

According to [*Popov, 2002*], the power index $\gamma$ can be obtained from the equation for the effective ionization frequency:

$$\gamma = \frac{\partial \ln(\nu_{ion} - \nu_{att})}{\partial \ln E} \qquad (2)$$

where $\nu_{ion}$ - $\nu_{att}$ is the difference between the ionization frequency and electron attachment frequency for a given sort of the gas. It depends only on the reduced electric field $E/N$ value.

The major results of the model [*Niemeyer et al., 1984; Wiesmann et al., 1986*] is that the plasma structures allow spatial scaling, thus they are self-similar structures or fractals. As shown in [*Satpathy, 1986*] the fractal dimention $D$ depends entirely on the dimensionality of space $d$ and the power index $\gamma$ in Eq. (1), $D = D(d, \gamma)$.

**Figure 1** reveals the power index as function of the reduced electric field $\gamma(E/N)$ for the air, here it is shown by the dashed line. It is computed for the $E/N$ range 200 - 1200 Td (1 Townsend is equal to $10^{-17}$ V/cm$^2$). The respective ionization and electron attachment frequencies $v_{ion}$(E/N) and $v_{att}$(E/N) were obtained by the Bolsig+ code [*Hagelaar et al.,* 2005].

In addition, Fig. 1 reveals the fractal dimension $D$ of the plasma structures formed in BJ, here it is shown by the solid line. The data $D(\gamma) = D(\gamma, d = 3)$ obtained in [*Satpathy,* 1986] for the 3D Laplace field was used. These data were determined by the model [*Niemeyer et al.,* 1984; *Wiesmann et al.,* 1986] from the solution of 3-D Laplas equation for the potential having the boundary conditions which consider the conducting channels. [*Satpathy,* 1986] computed the number of conducting channels N(R) which cross the sphere of radius R. Then the fractal dimension D(γ) was found using Eq. (5). The fractal dimension of the plasma structures vs. $E/N$ (Fig. 1) was found under the assumption that the maximum electric field in the streamer head is constant and equal to $E$.

The data [*Kuo et al.,* 2009] shows the reduced electric field in the streamer head inside the GBJ in the altitude range $h$ = 40 - 90 km. It was obtained using the ratio of the observed intensities of the optical bands N$_2$(2P) and N$_2$(1N). Found in the work [*Kuo et al.,* 2009] maximum reduced electric field was $E/N$ = 400 Td. Notice that due to the shift between the maximum electric field and maximum intensities of the N$_2$(2P) and N$_2$(1N) bands in the streamer head [*Naidis,* 2009; *Bonaventura et al.,* 2011] the real value $(E/N)_{max}$ in the streamers becomes about 1.4 times greater than the optical data gives. According to [*Kuo et al.,* 2009; *Bonaventura et al.,* 2011] we assume that in the streamer head where the branching probability is highest, the value $(E/N)_{max}$ reaches 500 - 600 Td (it is shown by the dotted area in Fig. 1).

As it follows from Fig. 1 for $E/N$ = 500 - 600 Td the fractal dimention of the branching plasma channels yields:

$$D = 2.05 \pm 0.04 \qquad (3)$$

Note that it significantly differes from $D = 1.37 \pm 0.13$ used by [*Pasko et al.*, 2000].

### 3. Average number of the streamers in blue jets

According to the fractal theory the full length $L$ of the streamers inside the surface having the radius R can be presented by [*Niemeyer et al.,* 1984; *Wiesmann et al.,* 1986]:

$$L \propto R^D \qquad (4)$$

While the number of streamers crossing this surface yields

$$N(R) = \frac{dL}{dR} = \left(\frac{R}{R_0}\right)^{D-1} , \quad (5)$$

where $R_0$ is the characteristic scale within which the streamer branches. The value $R_0$ cannot be obtained from the fractal theory and thus it should be provided as an input to the model.

Another important result from the fractal theory is that it describes the average inter-streamer distance $\Lambda$ as the function of $R$:

$$\Lambda(R) = \frac{d^2 L}{dR^2} \propto \left(\frac{R}{R_0}\right)^{D-2} \qquad (6)$$

Consider $D = 2$ (see Eq. (3)) we obtain that $\Lambda(R) \approx const$ i.e. the average inter-streamer distance changes only slightly as the distance R increases. Moreover, using Eq. (5) one can find the total number of the streamers in the Blue Jet, $N(R)$, at the height $h_{max}$. The latter is the maximum height where the optical steamers are the most distict, therefore they will be used in our analysis. Here $R = h_{max} - h_0$ and $h_0$ is the height at which the BJ starts, while $R_0$ is the characteristic scale of the leader head at the height $h_0$.

Knowing $N(R)$ one can find the average distance between the streamers

$$R_{av} = 2 \cdot R_{str} / \sqrt{N(R)} \qquad (7)$$

Here $R_{str}$ is the observed radius of the streamer corona at the height $h_{max}$. Furthermore the obtained value of $R_{av}$ was checked against the streamer radius observed at the height $h_{max}$.

## 4. Relation between the streamer speed and reduced diameter

In order to complete and test the fractal model we use the semi-empirical relations derived in [*Raizer et al.,* 1998; *Raizer et al.,* 2007]. The latter are based on studies of the laboratory streamers. The main input parameter for those relations is the voltage $\Delta U_t = U_t - U_0$ where $U_t$ and $U_0$ are the potential of the streamer head and of the external electric field respectively. The semi-empirical relations provides us with:

Field at the streamer head $E_m \approx 1.5 \cdot 10^4 (N / N_0)$ kV/m ;

Radius of the head $r_m \approx 3.3 \cdot 10^{-5} \Delta U_t (N_0 / N)$ m; $\Delta U_t$ [kV] ;

Speed of the positive streamer $v_s \approx 5.3 \cdot 10^4 \Delta U_t$ m/s; $\Delta U_t \geq 5$ kV ;

If the streamer radius is known the voltage $\Delta U_t$ and the streamer speed could be obtained from this model. On the other hand, knowing the streamer speed $V_{str}$ one can obtain $\Delta U_t$ and $r_m$. Note also that $\Delta U_t$ is unequivocally coupled to $E_m$ and $r_m$. For instance, for semi-spherical streamer head $\Delta U_t \approx E_m r_m$

Let us present the equations in terms of the reduced streamer diameter $d_{red} = 2\,r_m \cdot (N/N_0)$ and reduced electric field $E_m \cdot (N_0/N)$. We exclude $\Delta U_t$ from the equations for $r_m$ and $V_{str}$ and then obtain the following relation between the streamer speed and its reduced diameter

$$V_{str}(m/s) = 0.8 \cdot 10^6 \cdot d_{red} \,(mm) \qquad (8)$$

The solid line in Figure 5 shows the streamer speed *vs.* the reduced diameter computed with Eq. (8). The two dashed lines are adapted from [*Kanmae et al.,* 2012]. They were computed for the $E_s$ values 120 and 150 kV/cm, and validated by comparison with the observations of the Red Sprite streamers. Figure 5 shows that our semi-empirical model is in agreement with that of [*Kanmae et al.,* 2012].

### 5. Discussion

We begin with analysis of the BJ images [*Su et al.,* 2003; *Pasko et al.,* 2002; *Wescott et al.,* 2001] shown in Figs. 2-4. Listed in the top three rows of Table 1 are the hight $h_{st}$ where the BJ starts, and the maximum height $h_{max}$ where the optical streamers are most distinct, along with the radia of the streamer coronas $R_{str}$ measured at $h_{max}$.

Listed in Table 1 is also the relative air density at $h_{st}$ and $h_{max}$ altitude. It was obtained from the model of the earth atmosphere, and then used to compute the diameter of the leader head $R_0$. We assumed then that the diameter of the leader head $R_0$ is 1 cm at the ground, and it varies with the height inversely proportional to the air density.

Such height dependence is based on the following considerations. According to the modern understanding [*Bazelyan et al.,* 1998; *Liu et al.,* 2006] the value $R_0$ is determined by the characteristic rate of the electron attachment to the molecular oxygen. The latter process involves either two particles (e + $O_2 \rightarrow O^- + O$), or three particles (e + $O_2 + O_2 \rightarrow O_2^- + O_2$). At high altitude where BJs occur the dominant is the dissociative electron attachment, which is two particle reaction (see [*Liu et al.,* 2006]). Therefore we assume the linear dependence $R_0 \propto (N_0/N)$.

Theoretical estimates, based on the fractal model, show the total amount of the streamers $N_{tot}$ at $h_{max}$ and the average inter-streamer distance (see Eq. (7)).

As it shown in [*Milikh et al.,* 2016], in the streamer zone of a leader the strong inter-streamer interaction occurs when the streamer diameter is about the same as the distance between the streamers. Therefore we assume that the reduced streamer diameter is equal to the reduced inter-streamer distance.

Finally uing Eq. (8) we compute the streamer speed, which is in the range of $(0.3-2.0) \cdot 10^7$ m/s. Note that the experiments [*Wescott et al.,* 2001; *Pasko et al.,* 2002; *Su et al.,* 2003] lack fast speed optical detectors, thus the temporal resolution of the observed images was insufficient for accurate detection of the streamer speed. As mentioned by [*Pasko et al.,* 2002] "the estimated speed in the range $(1.9 - 2.2) \cdot 10^6$ m/s is therefore a lower bound on the actual speed, which was probably higher".

**Table 1**. Results of observations of Blue Jets and Gigantic Blue Jets adopted from [*Su et al.,* 2003; *Pasko et al.,* 2002; *Wescott et al.,* 2001] (the three top rows). Estimates based on the fractal model (rows 4-6 from the top). Reduced streamer diameters are shown in the row second from the bottom. The streamer speed is presented in the bottom row.

| Observables | [*Su et al.,* 2003] | [*Pasko et al.,* 2002] | [*Wescott et al.,*2001] |
|---|---|---|---|
| Height of streamers start $h_{st}$ | 37.5 km | 30 km | 20 km |
| Maximum height streamers reach $h_{max}$ | 73 km | 57 km | 40 km |
| Radius of the streamer corona $R_{str}$ at $h_{max}$ | 5.9 km | 5 km | 1.5 km |

| Estimates | [*Su et al.,* 2003] | [*Pasko et al.,* 2002] | [*Wescott et al.,*2001] |
|---|---|---|---|
| Relative air density $N(h=0)/N(h_{st})$ $N(h=0)/N(h_{max})$ | 204 24,500 | 66 2,450 | 13.8 306 |
| Total branches $N_{tot}$ at $h_{max}$ | 17,400 | 41,000 | 145,000 |
| Average inter-streamer distance, $R_{av}$ at $h_{max}$ | 89.4 m | 49.4 m | 7.8 m |
| Reduced streamer diameter | 3.7 mm | 20 mm | 25 mm |
| Streamer speed (from Eq. (8)) | $3 \cdot 10^6$ m/s | $1.6 \cdot 10^7$ m/s | $2 \cdot 10^7$ m/s |

6. **Conclusions.**

Multiple observations of Blue Jets (BJ) show that BJ emit a fan of streamers similar to a laboratory leader (streamer corona). In the exponential atmosphere those long streamers grow preferentially upward, producing a narrow cone confined by the aperture angle. In the presented paper the dendritic structure of streamers in corona discharge was treated by means of the fractal theory. The fractal dimension of the branching streamer channels was found.

Some BJ images available from the literature were analyzed and the average inter-streamer distance was estimated with the help of the fractal model. Based on our previous work we assumed that in the streamer zone of the leader the inter-streamer distance is about the same as the streamer diameter. Using this assumption and the semi-empirical model [*Raizer et al.,* 1998] we obtained the average streamer speed and compared it with the observations.

## Acknowledgments

This work was supported by the NSF grant 1226237.

**Figure Captions**

**Figure 1**. The fractal dimension D (solid line) and the power index gamma (the dashed line) versus reduced electric field E/N. Computed for the streamer heads.

**Figure 2.** Image of the gigantic blue jet [*Su et al.,* 2003]. The leader starts at $h_{st}$ = 37.5 km; the maximum altitude considered $h_{max}$ = 73 km; radius of the streamer corona $R_{str}$ = 5.9 km measured at $h_{max}$.

**Figure 3.** Image of the gigantic blue jet [*Pasko et al.,* 2002]. The leader starts at $h_{st}$ = 30 km; the maximum altitude considered $h_{max}$ = 57 km; radius of the streamer corona $R_{str}$ = 5 km measured at $h_{max}$.

**Figure 4.** Image of the blue jet [*Wescott et al.,* 2001]. The leader starts at $h_{st}$ = 20 km; the maximum altitude considered $h_{max}$ = 40 km; radius of the streamer corona $R_{str}$ = 1.5 km measured at $h_{max}$.

**Figure 5.** Streamer speed versus the reduced diameter. The solid line is computed using relation (7). The two dashed lines are adapted from [*Kanmae et al.,* 2012] where they were computed for $E_s$ = 120 and 150 kV/cm, and validated by comparison with the observations of the Red Sprite streamers.

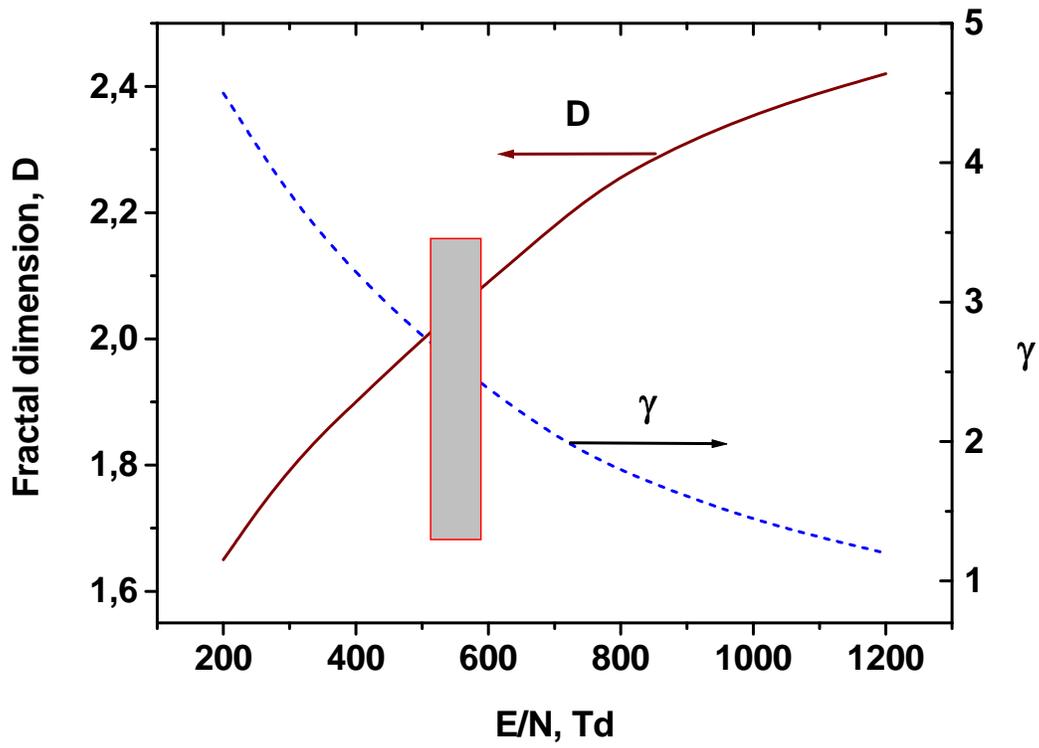

**Figure 1**. The fractal dimension D (solid line) and the power index gamma (the dashed line) versus reduced electric field E/N. Computed for the streamer heads.

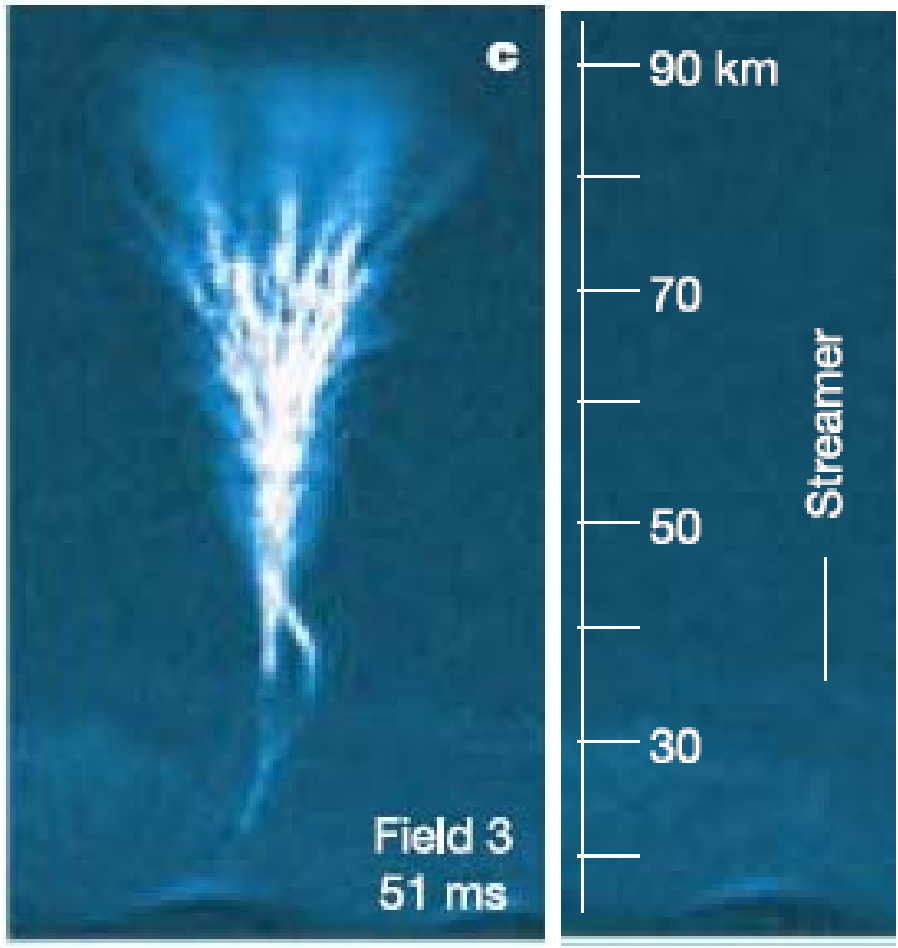

**Figure 2.** Image of the gigantic blue jet [*Su et al.,* 2003]. The leader starts at $h_{st}$ = 37.5 km; the maximum altitude considered $h_{max}$ = 73 km; radius of the streamer corona $R_{str}$ = 5.9 km measured at $h_{max}$.

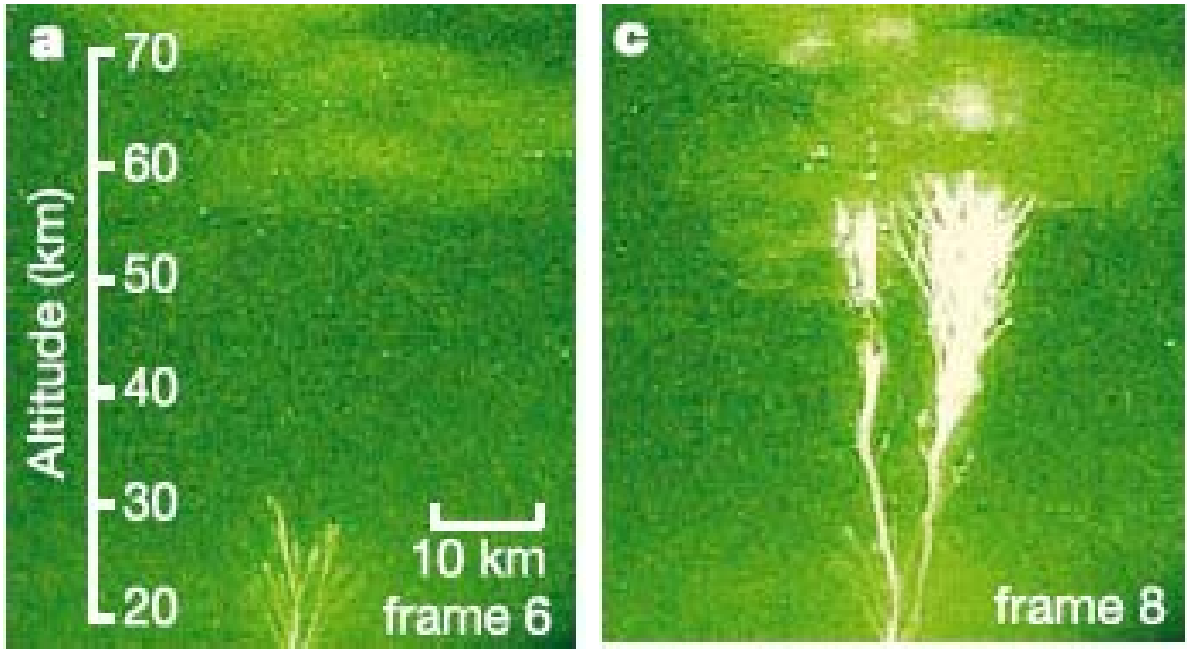

**Figure 3.** Image of the gigantic blue jet [*Pasko et al.,* 2002]. The leader starts at $h_{st}$ = 30 km; the maximum altitude considered $h_{max}$ = 57 km; radius of the streamer corona $R_{str}$ = 5 km measured at $h_{max}$.

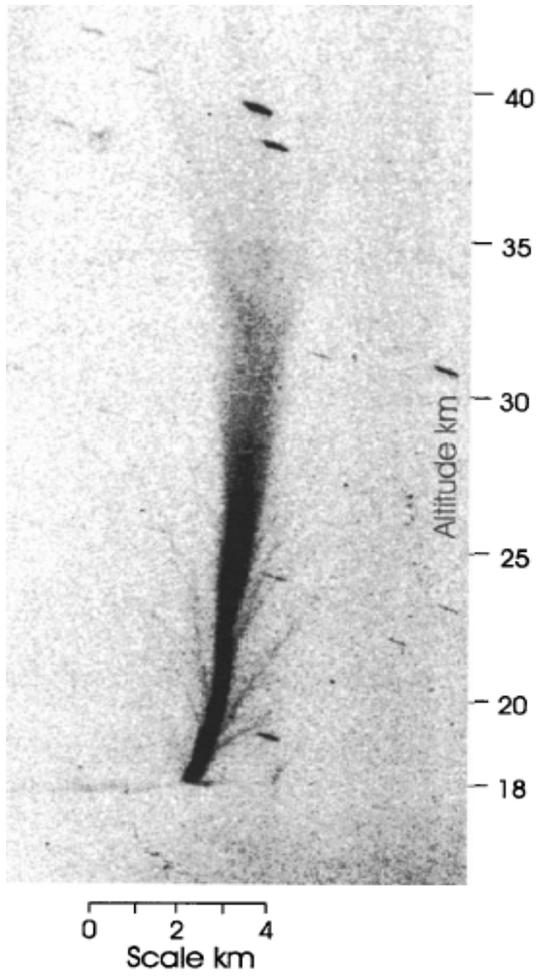

**Figure 4.** Image of the blue jet [*Wescott et al.,* 2001]. The leader starts at $h_{st}$ = 20 km; the maximum altitude considered $h_{max}$ = 40 km; radius of the streamer corona $R_{str}$ = 1.5 km measured at $h_{max}$.

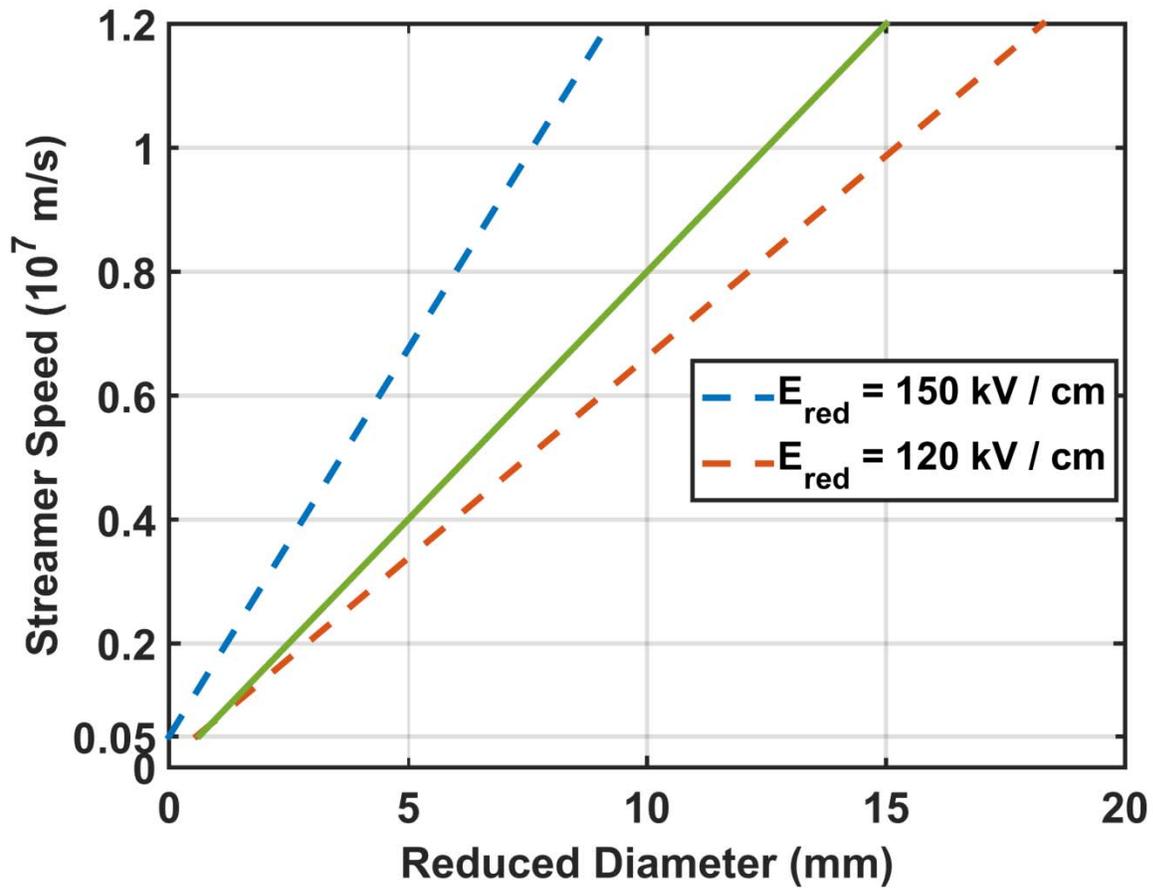

**Figure 5.** Streamer speed versus the reduced diameter. The solid line is computed using relation (8). The two dashed lines were computed for $E_{red}$ = 120 and 150 kV/cm, and validated by comparison with the observations of the Red Sprite streamers [*Kanmae et al.,* 2012].